\documentclass[
twocolumn,
]{ceurart}

\usepackage{listings}
\begin{document}

\copyrightyear{2022}
\copyrightclause{Copyright for this paper by its authors.
  Use permitted under Creative Commons License Attribution 4.0
  International (CC BY 4.0).}

\conference{IJCAI 2023 Workshop on Deepfake Audio Detection and Analysis (DADA 2023), August 19, 2023, Macao, S.A.R}

\title{ADD 2023: the Second Audio Deepfake Detection Challenge}

\author[1]{Jiangyan Yi}[%
email=jiangyan.yi@nlpr.ia.ac.cn
]
\cormark[1]
\address[1]{State Key Laboratory of Multimodal Artificial Intelligence Systems, Institute of Automation, Chinese Academy of Sciences, Beijing, China}

\author[2]{Jianhua Tao}[%
email=jhtao@tsinghua.edu.cn
]
\cormark[1]
\address[2]{Department of Automation, Tsinghua University, Beijing, China}

\author[1]{Ruibo Fu}[%
email=ruibo.fu@nlpr.ia.ac.cn
]

\author[1]{Xinrui Yan}[%
email=yanxinrui2021@ia.ac.cn
]
\fnmark[1]

\author[1]{Chenglong Wang}[%
email=chenglong.wang@nlpr.ia.ac.cn
]
\fnmark[1]

\author[1]{Tao Wang}[%
email=tao.wang@nlpr.ia.ac.cn
]
\fnmark[1]

\author[1]{Chu Yuan Zhang}[%
email=zhangchuyuan2021@ia.ac.cn
]
\fnmark[1]

\author[1]{Xiaohui Zhang}[%
]

\author[1]{Yan Zhao}[%
]

\author[1]{Yong Ren}[%
]

\author[1]{Le Xu}[%
]

\author[1]{Junzuo Zhou}[%
]

\author[1]{Hao Gu}[%
]

\author[1]{Zhengqi Wen}[%
]

\author[1]{Shan Liang}[%
]

\author[1]{Zheng Lian}[%
]

\author[1]{Shuai Nie}[%
]

\author[3,4]{Haizhou Li}[%
email=haizhou.li@u.nus.edu
]
\address[3]{Department of Electrical and Computer Engineering, National University of Singapore, Singapore}
\address[4]{The Chinese University of Hong Kong, Hong Kong}

\cortext[1]{Corresponding author.}
\fntext[1]{These authors contributed equally.}

\begin{abstract}
Audio deepfake detection is an emerging topic in the artificial intelligence community. The second Audio Deepfake Detection Challenge (ADD 2023) aims to spur researchers around the world to build new innovative technologies that can further accelerate and foster research on detecting and analyzing deepfake speech utterances. Different from previous challenges (e.g. ADD 2022), ADD 2023 focuses on surpassing the constraints of binary real/fake classification, and actually localizing the manipulated intervals in a partially fake speech as well as pinpointing the source responsible for generating any fake audio. Furthermore, ADD 2023 includes more rounds of evaluation for the fake audio game sub-challenge. The ADD 2023 challenge includes three subchallenges: audio fake game (FG), manipulation region location (RL) and deepfake algorithm recognition (AR). This paper describes the datasets, evaluation metrics, and protocols. Some findings are also reported in audio deepfake detection tasks.
\end{abstract}

\begin{keywords}
  Audio deepfake \sep
  fake detection \sep
  audio fake game \sep
  manipulation region location \sep
  deepfake algorithm recognition
\end{keywords}

\maketitle

\section{Introduction}

Over the last decades, the development of artificial intelligence has brought forth great improvements in speech synthesis ~\cite{tan2021survey,popov2021grad,kim2021conditional} and voice conversion~\cite{sisman2020overview,lorenzotrueba18_odyssey,Yi2020} technologies. The models are able to generate realistic and human-like speech. The technology nevertheless poses a serious threat to the society if someone misuses it~\cite{harwell2021remember}. Therefore, audio deepfake detection is an emerging topic of interest. An increasing number of efforts have been made to detect the deepfake audio recently~\cite{Wu2015ASVspoof,Kinnunen2017ASVspoof,Todisco2019ASVspoof,2021ASVspoof,yi2022add,Yi2021Half,Ma2021Continual}. 

A series of challenges, including Automatic Speaker Verification Spoofing and Countermeasures Challenge (ASVspoof 2021)~\cite{2021ASVspoof}, the First Audio Deepfake Detection Challenge (ADD 2022)~\cite{yi2022add} have played a critical role in fostering research on this area. The ASVspoof 2021 introduced a new task involving audio deepfake (DF) detection, accelerating progress in deepfake audio detection. To address more challenges in the real world, the ADD 2022~\footnote{http://addchallenge.cn/add2022}  included three Tracks: low-quality fake audio detection (LF), partially fake audio detection (PF) and audio fake game (FG). However, some limitations still existed in ADD 2022. The techniques used in the challenge focused more on performing binary classification between real and fake audio. In addition, there were limited rounds of evaluation for the FG Track. 

Moreover, there is also an interest in surpassing the constraints of binary real/fake classification, and actually localizing the manipulated intervals in a partially fake speech as well as pinpointing the source responsible for generating any fake audio. Therefore, we launched a second Audio Deepfake Detection Challenge (ADD 2023~\footnote{http://addchallenge.cn/add2023}) to spur researchers around the world to build new innovative technologies that can further accelerate and foster research on detecting and analysing deepfake utterances.

In the following sections, we describe the datasets and evaluation metrics designed for different subchallenges. Finally, we briefly report on the performance of the results submitted by the ADD 2023 participants to further explore the current state and future direction of real-world audio deepfake detection.

\section{Subchallenges}

The ADD 2023 challenge includes three subchallenges: audio fake game (FG) \cite{peng2021dfgc,peng2022dfgc}, manipulation region location (RL) and deepfake algorithm recognition (AR). The RL and AR subchallenges are new to ADD.

\textbf{Track  1. audio fake game (FG)}: 
Different from ADD 2022, ADD 2023 has two rounds of evaluations for the generation task and two rounds of evaluations for the detection task.

\textbf{Track  1.1 generation task (FG-G)}: aiming to generate fake audio that can fool the fake detection model of Track  1.2.

\textbf{Track  1.2 detection task (FG-D)}: attempting to detect fake utterances, especially the fake samples generated from Track  1.1.

\textbf{Track  2. manipulation region location (RL)}: focusing on locating the manipulated regions in a partially fake audio in which the original utterances are manipulated with real or generated audio \cite{Yi2021Half}.

\textbf{Track  3. deepfake algorithm recognition (AR)}: aiming to recognize the algorithms of the deepfake utterances, and the evaluation dataset includes samples from an unknown deepfake algorithm \cite{yan2022initial,yan2022system}.

\subsection{Training and dev sets}
The training and dev.\ sets of ADD 2023 contain four subsets, as summarized in Table 1. 

\begin{table}[h]
	\caption{Numbers of utterances in training and dev.\ sets of each task}
	\label{tab:train}
	\centering
	\begin{tabular}{cccccc}
		\toprule
	  \multirow{2}{*}{Subchallenges} & \multicolumn{2}{c}{Training} & \multicolumn{2}{c}{Dev.} \\ \cmidrule(l){2-5}
		&             \#Real &    \#Fake    & \#Real  &  \#Fake      \\ \midrule
		Track  1.2 &  3,012      & 24,072         &  2,307    & 26,017        \\
		Track  2 &   26,554   & 26,539            & 8,914     & 8,910  \\ 
  		Track  3 &   3,200   & 19,200            &   1,200   & 7,200  \\ \bottomrule
	\end{tabular}
\end{table}

\begin{table}[h]
	\caption{Overview of the numbers of utterances with different labels for the training set, dev.\ set and test set of Track 3. The unknown category is labeled 7.}
		\label{tab:train}
	\centering
	\begin{tabular}{cccccc}
		\toprule
	  Label  & Training &  Dev. & Test\\ \midrule
\#0   & 3,200 &  1,200  &9,512          \\
\#1  &  3,200  & 1,200   &10,474         \\ 
\#2  &   3,200   & 1,200   &7,169       \\ 
\#3  &   3,200   & 1,200    &10,461        \\
\#4  & 3,200   & 1,200     &10,391        \\
\#5    & 3,200   & 1,200     &10,507       \\
\#6   & 3,200   & 1,200   &10,507          \\
\#7   &-- & --  & 10,469\\
\bottomrule
	\end{tabular}
\end{table}

\textit{Track 1.1}: We use the AISHELL-3 \cite{Shi2020aishell3} corpus, which is a large-scale Chinese speech corpus containing over 88,000 utterances, composing 85 hours of speech. 

\textit{Track 1.2}: We use the same training and dev.\ sets as Track 3.2 of ADD 2022, including the real and fake utterances based on AISHELL-3.

\textit{Track 2}: The dataset  consists of real utterances and partially fake utterances. Fake utterances generated by manipulated the original genuine utterances with real or synthesized audios.

\textit{Track 3}: 
The training and dev.\ sets include 7 classes (1 real and 6 counterfeit) as shown in Table 2. The 7 categories are labeled 0, 1, 2, 3, 4, 5, 6. Fake audio taken from speech synthesized by different speech generation algorithms and tools.

\subsection{Test sets}
The test sets of ADD 2023 are more challenging compared to the previous one. The number of utterances in the four subsets are shown in Table 3.

\begin{table}[h]
	\caption{Numbers of utterances in the test sets of each task}
	\label{tab:test}
	\centering
	\begin{tabular}{ccccc}
		\toprule
		\multirow{2}{*}{Test} & \multicolumn{2}{c}{Track 1.2} & \multirow{2}{*}{Track 2} & \multirow{2}{*}{ Track 3} \\ \cmidrule(l){2-3}
		&        R1    &    R2     &  &  \\ \midrule
		\#Real & 80,000 & 87,500 & 
  20,000 & 10,507   \\
  \#Fake & 31,976 & 30,977 & 
  30,000 & 68,983   \\ 
  \bottomrule
	\end{tabular}
\end{table}

\textit{Track 1.1}: 
It consists of test sets for two rounds, with two speakers, one male and one female, randomly selected from the AISHELL3 dataset in each round. There are 499 text contents in the test set file, and the text content of each line corresponds to an audio file generated for each target speaker ID.

\textit{Track 1.2}: 
The real audio of the test set for two rounds consists of sources including AISHELL-1 \cite{8384449}, Thchs30 \cite{wang2015thchs}, etc. The fake audio consists of audio generated by using TTS and voice conversion techniques, and a portion of audio generated from the two rounds of track 1.1 submissions.

\textit{Track 2}: The test set includes unseen partially fake and real utterances. Additional noise addition and format conversions were done on this base.

\textit{Track 3}: 
The test set includes 8 classes (the 7 classes included in the training and dev.\ sets and unknown counterfeit class, as shown in Table 2). The unknown category data was synthesized by an unknown speech generation tool. 

\section{Evaluation metrics}
Track 1.1 aims to generate fake audio that can fool the detection models. Therefore, the deception success rate (DSR) \cite{yi2022add} is chosen as the metric. The goal of Track 1.2 is audio deepfake detection. So the weighted equal error rate (WEER) \cite{yi2022add} is used as the metric. To better evaluate the manipulation region location performance of Track 2, the final score is the weighted sum of sentence accuracy and segment F\textsubscript{1}-score \cite{Yi2021Half}. For Track 3, participants should recognize the known and unknown algorithms of the deepfake utterances. Therefore, we utilize the macro-average F\textsubscript{1}-score \cite{geng2020recent} in open set recognition. 

\subsection{Track 1.1 FG-G}
DSR reflects the degree to which the audio deepfake detection model is deceived by the generated utterances, and is defined as follows:
\begin{equation}
DSR =\frac{W}{A \times N}
\end{equation}
where $W$ is the count of wrong detection samples by all the detection models on the condition of reaching each own equal error rate (EER) \cite{Wu2015ASVspoof} performance, $A$ is the total number of evaluation samples, and $N$ is the number of detection models. For the first round, the DSR against the Track 1.2 submissions forms the totality of generation performance metric, where as in the second round, weighted consideration is also given to the DSR against the model we release, effectively:
\begin{gather}
WDSR = \gamma DSR_{R1}+ \delta WDSR_{R2}\\
WDSR_{R2} = \alpha DSR_{R2baseline} +\beta DSR_{R2}
\end{gather}
where $\gamma$=0.4, $\delta$=0.6, $\alpha$ =0.4 and $\beta$=0.6, and they represent the weights for DSR in our consideration. $\mathrm{DSR_{R1}}$ and $\mathrm{WDSR_{R2}}$ represent the generation performance metrics for the first and second rounds, respectively.
$\mathrm{DSR_{R2}}$ and $\mathrm{DSR_{R2baseline}}$ refers to the DSR achieved by using the 
synthesized speech submitted by the participants to attack the model submitted in track 1.2 and the detection baseline model~\footnote{https://github.com/asvspoof-challenge/2021/tree/main/DF/Baseline-RawNet2} provided by organizers, respectively.

\subsection{Track 1.2 FG-D}
The WEER is defined as:
\begin{equation}
WEER = \alpha EER_{R1} +\beta EER_{R2}
\end{equation}
where $\alpha$ =0.4 and $\beta$ =0.6, which represent the weights of $\mathrm{EER_{R1}}$ obtained in the first round and $\mathrm{EER_{R2}}$ obtained in the second round, respectively. The EER is defined and calculated in the same way as in ADD 2022.



\begin{table}
\caption{Description of detection baseline systems}
\label{tab:baselines}
\centering
\begin{tabular}{cccc}
\toprule
 ID & Model & Features & Task \\\midrule
S01 & GMM  &LFCC & Track 1.2 \\
S02 & LCNN  &LFCC & Track 1.2 \\
S03 &LCNN   & Wav2vec2 & Track 1.2  \\
S04 & LCNN  &LFCC & Track 2 \\
S05 & \makecell{ResNet\\(Softmax with threshold)} &LFCC & Track 3  \\
S06 & \makecell{ResNet\\(Openmax)} &LFCC &Track 3 \\
\bottomrule
\end{tabular}
\end{table}

\subsection{Track 2 RL}
For Track 2,  sentence accuracy measures the ability of the model to correctly distinguish between genuine and fake audio, and is defined as follows:
\begin{equation}
A_{sentence} =\frac{TP+TN}{TP+TN+FP+FN}
\end{equation}
where $TP$, $TN$, $FP$, and $FN$ denote the numbers of true positive, true negative, false positive, and false negative samples. Additionally, we use Segment Precision $\mathrm{P_{segment}}$, Segment Recall $\mathrm{R_{segment}}$, and Segment F\textsubscript{1}-score $\mathrm{F\textsubscript{1}_{segment}}$ to measure the ability of the model to correctly identify fake areas from fake audios, defined respectively as: 
\begin{equation}
P_{segment} =\frac{TP}{TP+FP}
\end{equation}
\vspace{-3pt}
\begin{equation}
R_{segment} =\frac{TP}{TP+FN}
\end{equation}
\vspace{-3pt}
\begin{equation}
F\textsubscript{1}_{segment} =\frac{2\times P \times R}{P+R}
\end{equation}
The final score is the weighted sum of Sentence Accuracy and Segment F\textsubscript{1}-score, as shown below.
\begin{equation}
Score = \alpha A_{sentence} + \beta  F\textsubscript{1}_{segment}
\end{equation}

where $\alpha$ =0.3 and $\beta$ =0.7, which represent the weights of $A_{sentence}$ and $F\textsubscript{1}_{segment}$.

\subsection{Track 3 AR}
For the algorithm recognition tasks in Track 3, we use the macro-average  F\textsubscript{1}-score, defined as:

\begin{equation}
P = \frac{1}{C} \sum \limits _{i=1} ^{C} \frac{TP_i}{TP_i+FP_i}
\end{equation}
\vspace{-3pt}
\begin{equation}
R = \frac{1}{C} \sum \limits _{i=1} ^{C}  \frac{TP_i}{TP_i+FN_i}
\end{equation}
\vspace{-3pt}
\begin{equation}
F_1 =\frac{2\times P \times R}{P+R}
\end{equation}
where $C$ denotes the number of known classes, $TP_i$, $FP_i$ and $FN_i$ denote the true positive, false positive, and false negative samples of class $i$ \cite{geng2020recent}. Note that while the formulae iterate only over known classes, $FP_i$ and $FN_i$ take unknown class samples into consideration.

\section{Challenge results}
ADD 2023 has challenge data requests from 145 teams from 12 countries. Participants submit task results and receive scores through the CodaLab website. In this section, we report on the detection baselines provided by the organizers and the results and analysis submitted by the participants.

\subsection{Detection baselines}

\begin{table*}
\centering
\caption{ADD 2023 Track 1.1 Rankings. The results of round 1 (R1) and round 2 (R2) were measured by DSR (\%) and the final evaluation was performed with WDSR (\%).}
\label{tab:track1.1}
\setlength{\tabcolsep}{1.3mm}{
\begin{tabular}{cccccc|cccccc}
\toprule
\# & ID & $\mathrm{DSR_{R1}}$ & $\mathrm{DSR_{R2}}$ & $\mathrm{DSR_{R2baseline}}$ & $\mathrm{WDSR}$ & \# &   ID & $\mathrm{DSR_{R1}}$ & $\mathrm{DSR_{R2}}$ & $\mathrm{DSR_{R2baseline}}$ & $\mathrm{WDSR}$  \\\midrule
1 & A01 & 37.91  & 49.60  & 49.80  & 44.97 & 9 & A09  & 0.00  & 35.55  & 24.85 & 18.76 \\
2 & A02 & 37.80  & 27.81  & 77.05  & 43.63 & 10 & A10  & 30.71  & 17.28  & 0.10 & 18.53 \\
3 & A03 & 43.20  & 51.58  & 23.45  & 41.48 & 11  & A11  & 23.58 &16.00 &6.71& 16.80 \\
4 & A04 & 33.16  & 36.25  & 51.30  & 38.63 & 12  & A12  & 41.72 &0.00 & 0.00 &16.69 \\
5 & A05 & 36.63  & 38.52  & 36.77  & 37.35 &13  & A13  & 40.12  &0.00 &0.00 & 16.05\\
6 & A06  & 38.14  & 36.66  & 9.32 & 30.69  & 14  & A14  & 0.00 &0.00 & 0.00 &0.00\\
7 & A07  & 39.68  & 22.79  & 25.45 & 30.18  & & Avg. & & &  &27.11 \\ 
8 & A08  & 34.83  & 29.03  & 5.51 & 25.71 \\
\bottomrule
\end{tabular}
\vspace{-8pt}
}
\end{table*}

\begin{table*}[h]
\caption{ADD 2023 Track 1.2 Rankings.
The results of round 1 (R1) and round 2 (R2) were measured by EER (\%) and the final evaluation was performed with WEER (\%).}
\label{tab:track3.2}
\centering
\setlength{\tabcolsep}{1mm}{
\begin{tabular}{ccccc|ccccc|ccccc}
\toprule
\# & ID & $\mathrm{EER_{R1}}$ & $\mathrm{EER_{R2}}$ & $\mathrm{WEER}$ & \# & ID & $\mathrm{EER_{R1}}$ & $\mathrm{EER_{R2}}$ & $\mathrm{WEER}$ & \# & ID & $\mathrm{EER_{R1}}$ & $\mathrm{EER_{R2}}$ & $\mathrm{WEER}$ \\\midrule
1 & B01 & 11.56  & 13.05  & 12.45  & 19 & B18 & 40.31  & 33.55  & 36.25  & 37 & B34 & 38.06  & 100.0  & 75.22 \\
2 & B02 & 21.11 & 15.82  & 17.93    & 20 & B19 & 54.71  & 28.46  & 38.96  & 38 & B35 & 38.06  & 100.0  & 75.47\\
3 & B03 & 23.44  & 21.26  & 22.13  & 21 & B20 & 32.90  & 45.04  & 40.18  & 39 & B36 & 41.34  & 100.00  & 76.53 \\
4 & B04 & 23.51  & 21.75  & 22.45  & 22 & B21 & 44.29  & 41.35  & 42.52  & 40 & B37 & 42.03  & 100.00  & 76.81 \\
5 & B05 & 24.06  & 22.59  & 23.17  & 23 & B22 & 41.91  & 43.76  & 43.02  & 41 & B38 & 42.38  & 100.00  & 76.95 \\
6 & B06 & 32.80  & 21.38  & 25.94  & 24 & B23 & 49.52  & 40.23  & 43.94  & 42 & B39 & 43.68  & 100.0  & 77.47 \\
7 & B07 & 32.73  & 24.94  & 28.05  & 25 & B24 & 42.83  & 45.24  & 44.27  & 43 & B40 & 44.50  & 100.0  & 77.80 \\
8 & B08 & 27.81  & 28.46  & 28.20  & 26 & B25 & 42.73  & 51.90  & 48.23  & 44 & B41 & 44.93  & 100.0  & 77.97 \\
9 &B09 & 32.91  & 25.50  & 28.46  & 27 & B26 & 44.61  & 51.48  & 48.73  & 45 & B42 & 47.04  & 100.0  & 78.81 \\
10 & B10 & 32.87  & 25.81  & 28.63  & 28 & \textbf{S01} & 36.90  & 63.80  & 53.04  & 46 & B43 & 47.04  & 100.0  & 78.81 \\
11 & B11 & 35.09  & 24.81  & 28.92  & 29 & B27 & 100.00  & 37.76  & 62.65  & 47 & B44 & 47.70  & 100.0  & 78.96 \\
12 & \textbf{S03} & 21.71  & 35.61  & 30.05  & 30 & B28 & 100.00  & 38.80  & 63.28  & 48 & B45 & 53.61  & 100.00  & 81.44 \\
13 & B12 & 32.37  & 29.93  & 30.90  & 31 & \textbf{S02} & 61.25  & 70.37  & 66.72  & 49 & B46 & 54.77  & 100.00  & 81.90 \\
14 & B13 & 33.34  & 30.50  & 31.63  & 32 & B29 & 25.94  & 100.00  & 70.37  & 50 & B47 & 54.91  & 100.0  & 81.96 \\
15 & B14 & 34.77  & 29.78  & 31.77  & 33 & B30 & 100.00  & 53.62  & 72.17  & 51 & B48 & 58.19  & 100.0  & 83.27 \\
16 & B15 & 38.75  & 30.08  & 33.54  & 34 & B31 & 35.33  & 100.00  & 74.13  & 52 & B49 & 65.00  & 100.0  & 86.00 \\
17 & B16 & 32.37  & 34.43  & 33.60  & 35 & B32 & 36.98  & 100.00  & 74.79  &   &Avg. &   &   &49.94 \\
18 & B17 & 38.08  & 34.20  & 35.75  & 36 & B33 & 100.00  & 58.04  & 74.82  &  &  &   &  &  \\
\bottomrule
\end{tabular}
\vspace{-8pt}
}
\end{table*}

ADD 2023 provides six baseline systems, which are described in summary as shown in Table 5. For the detection task of track 1.2, we present three different detection systems. The first system is a GMM-based system that operates on linear frequency cepstral coefficients (LFCCs) \cite{Sahi2015A} (baseline S01). The feature extraction of LFCC is the same as that of ASVspoof 2021, where the window length and shift are set to 30 ms and 15 ms. The second LFCC-LCNN system (baseline S02) operates on LFCC features with a light convolutional neural network (LCNN) \cite{Wu2020LCNN}. The frame length and shift are set to 20 ms and 10 ms. The back end is based on the LCNN reported in \cite{Wu2020LCNN}. The third system operates on wav2vec2 features with an LCNN (baseline S03). The wav2vec2 \cite{baevski2020wav2vec} pretrained model variant “wav2vec XLSR” is used as a pretrained feature extractor, which is trained on 56k hours of audio samples in 53 languages using additional linear transformations and a larger context network.

\begin{table}[h]
\caption{ADD 2023 Track 2 Rankings.
The results were performed by score (\%).}
\label{tab:track2}
\centering
\resizebox{0.5\textwidth}{!}{
\begin{tabular}{ccc|ccc|ccc}
\toprule
\# & ID &  Score & \# & ID &  Score & \# & ID &  Score \\\midrule
1 & C01 & 67.13  & 7 & C07 & 53.99  & 13 & \textbf{S04} & 42.25 \\
2 & C02 & 62.49  & 8 & C08 & 50.86  & 14 & C13 & 42.11 \\
3 & C03 & 62.02  & 9 & C09 & 48.55  & 15 & C14 & 38.74 \\
4 & C04 & 59.62  & 10 & C10 & 45.39  & 16 & C15 & 27.57 \\
5 & C05 & 59.12  & 11 & C11 & 44.65  & 17  & C16 & 18.80 \\
6 & C06 & 56.63  & 12 & C12 & 43.50  &   & Avg. & 48.82 \\
\bottomrule
\end{tabular}
}
\end{table}

\begin{table}[h]
\caption{ADD 2023 Track 3 Rankings.
The results were performed by F\textsubscript{1}-score (\%).}
\label{tab:track}
\centering
\setlength{\tabcolsep}{1.3mm}{
\begin{tabular}{ccc|ccc|ccc}
\toprule
\# & ID &  F\textsubscript{1} & \# & ID &  F\textsubscript{1} & \# & ID &  F\textsubscript{1} \\\midrule
1 & D01 & 89.63  & 6 & D06 & 73.50  & 11 & \textbf{S05} & 53.50 \\
2 & D02 & 83.12  & 7 & D07 & 72.05  & 12 & D10 & 21.10 \\
3 & D03 & 75.41  & 8 & D08 & 68.15  & 13 & D11 & 11.73 \\
4 & D04 & 73.55  & 9 & D09 & 63.78  &  & Avg. & 62.87  \\
5 & D05 & 73.52  & 10 & \textbf{S06} & 54.16  &  \\
\bottomrule
\end{tabular}
}
\end{table}

For the detection task of track 2, the front-end LFCC feature extraction settings of the baseline system S04 are the same as those of S02. For back-end model architecture, we remove all pooling layers from the conventional LCNN to ensure the output size aligns with the segment label. For the recognition task of track 3, we introduce two different recognition systems. Both baselines are LFCC-ResNet based systems. The LFCC were extracted similar to baseline system S01. The model structure of ResNet based on \cite{he2016deep} was adopted from ResNet-18. Baseline S05 used a simple method-softmax with threshold, a thresholding procedure on probability to identify whether a speech signal is generated by a known or an unknown deepfake algorithm. Baseline S06 used the traditional open-set recognition method of OpenMax \cite{bendale2016towards}. 

\subsection{Results and analysis}
The four tracks of ADD 2023 have all received sufficient submissions, and the summary data of the rankings are shown in Tables 5, 6, 7, and 8. The ID number of each participating team is determined by their ranking order. 

Track 2 and 3 are the first subchallenge of fake region location and algorithm recognition in the field of deepfake audio detection.
For track 1.1, we received 14 submissions. The average WDSR of all submissions was 27.11\%, and the two-round combined performance of the best team was 44.97\%. 
Track 1.2 received a total of 49 submissions, with 11 WEER below the best baseline S01, and the best team had a WEER of 12.45\%. The average WEER of all submissions was 49.94\%.
For Track 2, 11 teams scored higher than the baseline S04, with the highest score of 67.13\%. The average score of the 16 submissions was 48.82\%. The results show that it is challenging for fake region location..

For Track 3,  Nine teams performed better than the baseline systems S06 and S07. Although the best team achieved an F\textsubscript{1}-score of 89.63\%, the average F\textsubscript{1}-score of Track 3 is still low. We hope that the challenge data and evaluation results of track 3 will further promote researchers to explore new deepfake audio algorithm recognition methods.

\section{Conclusions}

This paper provides an overview of the ADD 2023 Challenge, which consists of four distinct subchallenges. In order to better simulate real-world challenges, the challenge introduces two new tasks and more difficult datasets.  The results indicate that the fake region location task and the  algorithm recognition task are still challenging, especially for fake region location track. The solutions of participants and further analysis will be presented at the ADD 2023 workshop. In future competitions, we plan to optimize the datasets and competition rules, aiming to promote more advanced research in the deepfake audio community.

\begin{acknowledgments}
This work is supported by  the National Natural Science Foundation of China (NSFC) (No.\ 61831022, No.\ U21B2010, No.\ 62101553, No.\ 61971419, No.\ 62006223, No.\ 62276259, No.\ 62201572, No.\ 62206278), Beijing Municipal Science\&Technology Commission, Administrative Commission of Zhongguancun Science Park No.\ Z211100004821013, Open Research Projects of Zhejiang Lab (NO.\ 2021KH0AB06). Thanks to AISHELL ~\footnote{https://www.aishelltech.com}  for providing the open source dataset for this challenge.  
\end{acknowledgments}

\bibliography{sample-ceur}

\appendix

\end{document}